\documentclass[aps,eqsecnum,pra]{revtex4}
\usepackage{graphics,graphicx}
\usepackage{amsmath}
\usepackage{amssymb,latexsym,mathrsfs}
\usepackage{hyperref}

\def\bea{\begin{eqnarray}}
\def\eea{\end{eqnarray}}
\def\ba{\begin{array}}
\def\ea{\end{array}}

\def\beq{\begin{equation}}
\def\eeq{\end{equation}}

\begin{document}

\title{Preservation of coherence for a Two-Level atom tunneling through a squeezed vacuum with finite bandwidth}

\author{Samyadeb Bhattacharya$^{1}$ \footnote{sbh.phys@gmail.com}, Sisir Roy$^{2} $ \footnote{sisir@isical.ac.in}}
\affiliation{$^{1,2}$Physics and Applied Mathematics Unit, Indian Statistical Institute, 203 B.T. Road, Kolkata 700 108, India \\}

\vspace{2cm}
\begin{abstract}

\vspace{1cm}

\noindent Tunneling of a two-state particle through a squeezed vacuum is considered. It has been shown that repetitive measurement or interaction with the external field can preserve the coherence. Moreover, the coherence time in terms of the squeezing parameters has been calculated. A specific condition is derived, under which the coherence is sustainable.

\vspace{2cm}


\vspace{1cm}
\end{abstract}

\vspace{1cm}

\maketitle

\section{Introduction}
Starting from quantum computation to various other aspects, the ability to preserve coherence in quantum systems is of fundamental importance with many useful consequences. But in practical situations, quantum mechanical systems are extremely vulnerable to environmental interactions, leading to the loss of coherence in a very short period of time. The process of this breaking up of quantum superposition is known as the phenomena of Decoherence \cite{1}. Developing techniques to preserve quantum coherence is an emergent area of research nowadays from both theoretical and experimental point of view. The spin echo and multiple pulse techniques in NMR \cite{2,3}, dynamical decoupling methods for open quantum systems \cite{4,5} are the examples of some versatile tools for controlling decoherence. Environment, in the form of some external fields plays the role of reservoir in open quantum systems. Sometimes squeezed vacuums can also assume the role of the reservoir \cite{6,7,8}. Though the properties of such reservoirs are different. It is a quite well known fact that, squeezed vacuum can have considerable effects on quantum dissipative processes \cite{9}. Particularly, when a two-state atomic system interacts with a broadband squeezed vacuum, the transverse polarization quadratures exhibit decay processes, though different from the usual quantum decays \cite{6}. In this work, we have considered a system of two-level atom tunneling through a squeezed vacuum with finite bandwidth. Our goal is to investigate the possibilities of sustainable quantum coherence within the period of dwelling through the vacuum region. \\ In the first part, we will calculate the coherence time and the dwell time \cite{10} for the system and compare the two timescales. The ratio of these two timescales is considered as a measure of sustainable coherence within the period of tunneling through the vacuum. If the coherence time is longer than the dwell time (ie. the ratio is greater than unity), then it can be said that quantum coherence is sustainable for at least the period of tunneling through the barrier. From our result, it will follow that repetitive measurement is one particular condition, under which we can achieve sustainable coherence. In the second part of our work, we will consider the master equation for the system coupled to the squeezed bath, from which we will derive the decay parameter for the particle under steady state scenario. Considering this decay dynamics, we will then formulate the coherence time for the particle tunneling through the vacuum. We will see that this timescale is dependent on the system frequency as well as a number of bath parameters, leading to a specific condition under which the coherence of the system is sustainable. In Section II, we will discuss the aspect of coherence control through repetitive measurement and see that how Zeno dynamics can play an important role in sustaining the quantum coherence of the system. Then in Section III, we will consider the master equation approach for a two-state system tunneling through a squeezed vacuum and derive the coherence time. After that we will conclude with some possible implications.

\section{Preservation of coherence by repetitive measurement }
Let us now start with the concept of degree of coherence, which is a certain measure of the amount of coherence for the system in concern. Since in this work, we are dealing only with the time evolution of the system, so we will consider the  degree of temporal coherence, which is given by the expression

\beq\label{1}
g(\tau)=\frac{G(\tau)}{G(0)}
\eeq

where $G(\tau)$ is the coherence function given by

\beq\label{2}
G(\tau)=\lim_{T\to \infty} \frac{1}{2T}\int^{T}_{-T} U_s^{*}(t)U_s(t+\tau) dt
\eeq
where $U_s(t)$ is the time evolution operator of the system, given by

\beq\label{3}
U_s(t)=\exp[i\alpha t-\Gamma t]
\eeq
$\Gamma$ is the decay parameter, which we will evaluate in later course. Using the expression of $U_s(t)$, we can get from eqn (\ref{2})

\beq\label{4}
G(\tau)=\lim_{T\to \infty}  \frac{\sinh(2\Gamma T)}{2\Gamma T}\exp[i\alpha \tau-\Gamma \tau]
\eeq
So the degree of coherence is found to be

\beq\label{5}
g(\tau)= \exp[i\alpha \tau-\Gamma \tau]
\eeq

The coherence time for the system can be calculated as

\beq\label{6}
\tau_C= \int_{0}^{\infty} |g(\tau)|^2 d\tau = \frac{1}{2\Gamma}
\eeq

Now let us derive the expression of dwell time for our concerning system in the framework of weak measurement. Weak measurement of a certain operator \cite{12,13,14} is the process of measurement which is done on an ensemble of pre and post selected states, keeping the interaction between measurement device and the system sufficiently weak. In our case, measurement means interaction with the external electromagnetic field, which is playing the role of reservoir. This external field is acting as the measurement device. We are keeping the interaction with this field "weak", so a single measurement (ie. interaction) of certain operation by the field do not give any significant result. But an ensemble average of many such interactions will give us a certain meaningful result. So in that sense we are allowed to call it weak measurement. Because of the interactions with the environmental modes, majority of the states in the respective Hilbert space become highly unstable. For that reason, after a short period of time, the system decays into a particular state, which is a mixture of some simple pointer states. In a previous work \cite{11}, we have derived the expression of the weak value of dwell time in dissipative system on the basis of the procedure developed by Aharonov et.al. \cite{15}. We will mainly follow the same framework.\\
Dwell time is defined as the time interval, within which the particle resides within the barrier region. We can construct a certain operator determining whether the particle is within the barrier region or not.
\beq\label{7}
\Theta_{(0,L)} = \Theta(x)-\Theta(x-L)
\eeq
where $\Theta$ is a heaviside function and $L$ is the width of the vacuum (the barrier).

\beq\label{8}
\Theta_{(0,L)}= \left \{  \begin{array}{ll}
                 1 & \mbox{if}~~~ 0<x<L\\
                 0 & \mbox{otherwise}
                 \end{array} \right.
\eeq

For an observable $A$, dividing the total measurement into many short intervals ($\delta t$), the weak value of the operator $A $($ A\simeq \sum_{j=-\infty}^{\infty} A_j  $) , over an ensemble of the pre and post selected states $|\psi_i\rangle$ and $|\psi_f\rangle$ respectively, is given by \cite{15}

\beq\label{9}
<A_j>^w=C\Delta t \frac{\langle\psi_f(j\Delta t)|A|\psi_i(j\Delta t)\rangle}{\langle\psi_f(j\Delta t)|\psi_i(j\Delta t)\rangle}
\eeq
where $C$ is some arbitrary constant. We can set it as $C=\frac{1}{\delta t}$. For $\delta t \rightarrow 0$ and $A=\Theta_{(0,L)}$, replacing the summation by integration, we get

\beq\label{10}
<\tau>^w= \frac{\int_{-\infty}^{\infty}dt\int_0^L \psi_f^{*}(x,t) \psi_i(x,t)dx}{\int_{-\infty}^{\infty} \psi_f^{*}(x,0) \psi_i(x,0)dx}
\eeq

Let us initially define our system Hamiltonian as

\beq\label{11}
H_s=\frac{1}{2}\hbar \Omega \sigma_z
\eeq

So the time evolution operator looks like

\beq\label{12}
U(t)= \left (  \begin{array}{ll}
                 e^{i\Omega t/2} & 0\\
                 0 & e^{-i\Omega t/2}
                 \end{array} \right)
\eeq

Let us define the pre-selected state polarized in the x-direction as
\beq\label{13}
|\psi_i\rangle= \frac{1}{\sqrt{2}}\left (\begin{array}{ll}
                         1\\
                         1
                 \end{array} \right)
\eeq

The projection operator onto this pre-selected state
\beq\label{14}
P_{+}= \frac{1}{\sqrt{2}}\left (  \begin{array}{ll}
                 1 & 1\\
                 1 & 1
                 \end{array} \right)
\eeq

Now let us consider the decay of this pre-selected state, due to the interaction with the environmental squeezed modes \cite{16}. Consider the excited state $E_n$ to satisfy the relation

\beq\label{15}
E_n-E_0=n\Delta E,~~~~~-N\leq n\leq N
\eeq
The chosen excited states are equispaced and distributed symmetrically about the excited state of the reference atom, which is taken as $n=0$. According to Davies \cite{16}, the time evolution operator for the relevant sub-space can be depicted by a $(2N+1)\times(2N+1)$ dimensional matrix. The components of the matrix
\beq\label{16a}
U_{00}=e^{-\Gamma t}
\eeq

\beq\label{16b}
U_{n0}=iH_s\left[\frac{e^{-\Gamma t+in\Delta Et}-1}{\Gamma -in\Delta E}\right ]
\eeq
Without going into further detail, which can be found in references \cite{11,16}, we say that in this procedure the weak value of the projection operator defined in (\ref{14}), can be found as
\beq\label{17}
P_w=e^{-\Gamma(t-t_i)} \left[\frac{1-e^{-\Gamma(t_f-t)+ik\Delta E(t_f-t)}}{1-e^{-\Gamma(t_f-t_i)+ik\Delta E(t_f-t_i)}}\right]
\eeq
where the post selected final state is
\beq\label{18}
|\psi_f\rangle=|\psi_k\rangle
\eeq

The initial pre-selected state is that with energy $E_0$. Now for the specific choice of $E_k=E_0$, (\ref{17}) is modified as
\beq\label{19}
P_w=e^{-\Gamma(t-t_i)} \left[\frac{1-e^{-\Gamma(t_f-t)}}{1-e^{-\Gamma(t_f-t_i)}}\right]
\eeq
where $t_i$ and $t_f$ are respectively the instant of initial and final interaction with the field. So $t_f-t_i=\tau_m$ can be considered as sort of a measurement time. As we have mentioned earlier, the pre-selected state is that with energy $E_0$. Now, if the post selected state is also chosen as $E_0$ state, then eqn (\ref{19}) will give us the weak value of the survival probability. Here we interpret the time integral of this survival probability as the weak value of dwell time. This understanding conforms with the interpretation of dwell time as given in \cite{17,18,19,20}. There the barrier is understood as a sort of capacitive region which accumulates and scatters the energy falling upon it in the form of incident wave. Dwell time is the time delay between the accumulation and scattering of the energy. Following this argument, we interpret the projection operator $P_{+}$ as the operator $\Theta_{(0,L)}$ and find the weak value of dwell time to be
\beq\label{20}
\begin{array}{ll}
\tau_D^w=\int_{t_i}^{t_f} e^{-\Gamma(t-t_i)} \left[\frac{1-e^{-\Gamma(t_f-t)}}{1-e^{-\Gamma(t_f-t_i)}}\right]dt \\
~~~~=\frac{1}{\Gamma}\left[1-\frac{\Gamma\tau_m}{e^{\Gamma\tau_m}-1}\right]
\end{array}
\eeq
The measurement time $\tau_m$ is the time delay between two successive interactions. If we assume the measurement interaction is frequent enough, so that $\tau_m\ll 1/\Gamma$, then the dwell time reduces to be
\beq\label{21}
\tau_D^w=\frac{1}{2/\tau_m+\Gamma}
\eeq

Now comparing eqn (\ref{6}) and (\ref{21}), we find the ratio of coherence time and dwell time as
\beq\label{22}
\frac{\tau_C}{\tau_D^w}= \frac{1}{2} + \frac{1}{\tau_m \Gamma}
\eeq

From this equation, we infer that with the condition $\tau_m\Gamma\ll 1$, the ratio $\tau_C/\tau_D^w$ becomes very large. ie. the coherence time exceeds the dwell time considerably. If the decay parameter $\Gamma$ is very small, the ratio increases. This is expected. Because small decay parameter means the environmental interaction is not considerable; ie. the system is more or less isolated, for which the quantum coherence is preserved. It's the interaction with the environment, that drives the loss of quantum coherence, making the system more and more classical. But more interestingly on the other hand, decreasing measurement time ($\tau_m$) also makes the ratio to increase. ie. if we make the measurement more and more frequent, we can make the coherence more and more sustainable. This is a very interesting inference. We know for a fact that in some cases frequent measurement over a decaying quantum state slows down the decay making the system to freeze. This phenomena is known as ``Quantum Zeno Effect" \cite{21}. Frequent measurement forces the system to evolve in a reduced subspace of the total Hilbert space \cite{22}. These reduced zeno subspaces act more like an isolated space, within which the ``quantumness" of the concerning system can be preserved. From eqn (\ref{22}), we can explicitly show that frequent measurement in the form of field interaction can preserve the coherence of the system for a considerably longer period of time. \\

\section{Condition for sustainable coherence for a two-state particle tunneling through squeezed vacuum}
Let us now move on to the second part of our work, where we have considered the master equation for the two-state system coupled to the squeezed vacuum and calculate the coherence time in terms of the system and bath parameters. One of the fundamental properties of squeezed states is that they reduces quantum fluctuation. Squeezing has been studied by many researchers over the past years \cite{7,24,25}. One of the first papers discussing squeezed vacuum was published by Gardiner \cite{6}. We will follow the work of Tana\'{s} \cite{27}, where the problem of two level atom in squeezed vacuum is dealt in the master equation approach. Let us define the total Hamiltonian of the system and reservoir in the usual way as
\beq\label{23}
H_T=H_s+H_{em}+H_{int}
\eeq

where $H_s,H_{em}$ and $H_{int}$ are the system, reservoir and interaction Hamiltonian respectively. They are described by \cite{6}
\beq\label{24}
\begin{array}{ll}
H_s=\frac{1}{2}\hbar\Omega\sigma_z+\frac{1}{2}\hbar\xi\sigma_{x}\\
H_{em}=\hbar\int_0^{\infty} d\omega \omega b^{\dag}(\omega)b(\omega)\\
H_{int}=i\hbar\int_0^{\infty} K(\omega)[b^{\dag}(\omega)\sigma^{-}-b(\omega)\sigma^{+}]d\omega
\end{array}
\eeq

where $b^{\dag}(\omega)$ and $b(\omega)$ are respectively the creation and annihilation operators for the reservoir Hamiltonian, which is assumed as a collection of harmonic oscillators. The $\sigma_x$ part is the perturbation part with small complex energy $\xi$ corresponding attenuation. $K(\omega)$ is the coupling parameters for the interaction Hamiltonian. The correlation functions for $b^{\dag}(\omega)$ and $b(\omega)$ can be given as \cite{6}
\beq\label{25}
\begin{array}{ll}
\langle b(t)\rangle=\langle b^{\dag}(t)\rangle=0, \langle b^{\dag}(t)b(t')\rangle=N\delta(t-t')\\
\langle b(t)b^{\dag}(t')\rangle=(N+1)\delta(t-t')\\
\langle b(t)b(t')\rangle=M e^{-i\Omega(t+t')}\delta(t-t')\\
\langle b^{\dag}(t)b^{\dag}(t')\rangle=M^{*} e^{i\Omega(t+t')}\delta(t-t')
\end{array}
\eeq

where $N(\omega)$ and $|M(\omega)|$ are given by

\beq\label{26}
\begin{array}{ll}
N(\omega)=\frac{\lambda^2-\mu^2}{4}\left[\frac{1}{x^2+\mu^2}-\frac{1}{x^2+\lambda^2}\right]\\
|M(\omega)|=\frac{\lambda^2-\mu^2}{4}\left[\frac{1}{x^2+\mu^2}+\frac{1}{x^2+\lambda^2}\right]
\end{array}
\eeq
where $x=\omega-\omega_L$ is the shift from laser frequency $\omega_L$. $\lambda$ and $\mu$ are related to the cavity damping rate $\gamma$ and the real amplification constant $\epsilon$

\beq\label{27}
\lambda=\frac{1}{2}\gamma +\epsilon ,~~~~ \mu=\frac{1}{2}\gamma -\epsilon
\eeq

and $M=|M|e^{i\phi}$, where $\phi$ is the phase of squeezing. The cavity dissipation parameter $(\gamma)$ can be related to the coupling parameter $(K(\omega))$ as $\gamma=2\pi K(\Omega)^2$ \cite{6}. So the resulting master equation can be taken of the form \cite{27}
\beq\label{28}
\begin{array}{ll}
\dot{\rho}=\frac{1}{2}i\gamma\delta[\sigma_z,\rho]\\
+\frac{1}{2}\gamma\widetilde{N}(2\sigma_{+}\rho\sigma_{-}-\sigma_{-}\sigma_{+}\rho-\rho\sigma_{-}\sigma_{+})\\
+\frac{1}{2}\gamma(\widetilde{N}+1) (2\sigma_{-}\rho\sigma_{+} -\sigma_{+}\sigma_{-}\rho-\rho\sigma_{+}\sigma_{-})\\
-\gamma\widetilde{M}\sigma_{+}\rho\sigma_{+}-\gamma\widetilde{M}^{*}\sigma_{-}\rho\sigma_{-}-\frac{1}{2}i\Omega[\sigma_{+}+\sigma_{-},\rho]\\
+\frac{1}{4}i(\beta[\sigma_{+},[\sigma_z,\rho]]-\beta^{*}[\sigma_{-},[\sigma_{z},\rho]])

\end{array}
\eeq
where
\beq\label{29}
\widetilde{N}=N(\omega_L+\Omega')+\frac{1}{2}(1-\widetilde{\Delta}^2)Re\Upsilon_{-}
\eeq

\beq\label{30}
\widetilde{M}=M(\omega_L+\Omega')-\frac{1}{2}(1-\widetilde{\Delta}^2)\Upsilon_{-}+i\widetilde{\Delta}\delta_Me^{i\phi}
\eeq

\beq\label{30a}
\begin{array}{ll}
\mathbf{{\Upsilon_{-}=N(\omega_L)-N(\omega_L+\Omega')}}\\
~~~~~~-[|M(\omega_L)|-|M(\omega_L+\Omega')|]e^{i\phi}
\end{array}
\eeq

\beq\label{31}
\delta=\frac{\Delta}{\gamma}-\frac{1}{2}(1-\widetilde{\Delta}^2)Im\Upsilon_{-}+\widetilde{\Delta}\delta_N
\eeq

\beq\label{32}
\beta=\gamma\widetilde{\Omega}\left[\delta_N+\delta_M e^{i\phi}-i\widetilde{\Delta}\Upsilon_{-}\right]
\eeq

\beq\label{33}
\Omega'=\sqrt{\Omega^2+\Delta^2},~~~~\widetilde{\Omega}=\frac{\Omega}{\Omega'},~~~~\widetilde{\Delta}=\frac{\Delta}{\Omega'}
\eeq
$\Omega$ is the Rabi frequency. $\Delta=\omega_L-\omega_A$ is the detuning of the laser field, where $\omega_L$ and $\omega_A$ are the laser frequency and atomic frequency respectively. $\delta_N$ and $\delta_M$ are the squeezing induced shifts, which depend on $N(\omega)$ and $|M(\omega)|$ respectively.\\
From the master equation, it can be shown that \cite{27}

\beq\label{34}
\begin{array}{ll}
\langle\dot{\sigma}_{-}\rangle=-\gamma(\frac{1}{2}+\widetilde{N}-i\delta)\langle\sigma_{-}\rangle
-\gamma\widetilde{M}\langle\sigma_{+}\rangle+\frac{i}{2}\Omega\langle\sigma_z\rangle\\
\langle\dot{\sigma}_z\rangle=i(\Omega+\beta^{*})\langle\sigma_{-}\rangle
-i(\Omega+\beta)\langle\sigma_{+}\rangle\\
~~~~~~~~~~~~-\gamma(1+2\widetilde{N})\langle\sigma_z\rangle-\gamma

\end{array}
\eeq
The equation for $\sigma_{+}$ is nothing but the hermitian conjugate of the equation for $\sigma_{-}$. The hermitian operators $\sigma_x$ and $\sigma_y$ can be defined as

\beq\label{35}
\begin{array}{ll}
\sigma_x=\frac{1}{2}(\sigma_{-}+\sigma_{+})\\
\mathbf{\sigma_y=\frac{i}{2}(\sigma_{-}-\sigma_{+})}
\end{array}
\eeq
Solving the equations defined in (\ref{34}), we get

\beq\label{36}
\langle\sigma_z\rangle_{ss}=-\gamma\frac{\gamma^2\left(\frac{1}{4}+\widetilde{N}(\widetilde{N}+1)-|\widetilde{M}|^2+\delta^2\right)}{d}
\eeq

\beq\label{37}
\langle\sigma_{+}\rangle_{ss}=\langle\sigma_{-}\rangle^{*}_{ss}=i\frac{\Omega}{2d}\gamma^2\left(\frac{1}{2}+\widetilde{N}+\widetilde{M}^{*}-i\delta\right)
\eeq

where
\beq\label{38}
\begin{array}{ll}
d=\gamma^3(1+2\widetilde{N})\left(\frac{1}{4}+\widetilde{N}(\widetilde{N}+1)-|\widetilde{M}|^2+\delta^2\right) \\
+\gamma\Omega\left[\left(\frac{1}{2}+\widetilde{N} +Re\widetilde{M}\right)(\Omega+Re\beta)+Im\beta(Im\widetilde{M}+\delta)\right]
\end{array}
\eeq
The time evolution operator for the system in the steady state scenario can be expressed after tracing out over the bath variables

\beq\label{39}
U_{s}=\exp\left[-\frac{i}{2}\left(\Omega\langle\sigma_z\rangle_{ss}+\xi\langle\sigma_{x}\rangle_{ss}\right)t\right]
\eeq

putting the values of $\langle\sigma_z\rangle_{ss}$ and $\langle\sigma_{x}\rangle_{ss}$ using (\ref{35}),(\ref{36}) and (\ref{37}), in equation (\ref{39}), we get

\beq\label{40}
\begin{array}{ll}
U_{s}=\exp[-\frac{i}{2}(-\Omega\gamma\frac{\gamma^2(\frac{1}{4}\widetilde{N}(\widetilde{N}+1)-|\widetilde{M}|^2+\delta^2)}{d}\\
~~~~~~~~~~+i\xi\frac{\Omega\gamma^2}{4d}(-2i\delta-2Im\widetilde{M}))t]\\
~~~=\exp[\frac{i}{2}(\Omega\gamma\frac{\gamma^2(\frac{1}{4}\widetilde{N}(\widetilde{N}+1)-|\widetilde{M}|^2+\delta^2)}{d}-\xi\frac{\Omega\gamma^2}{2d}\delta)t\\
~~~~~~~~~~~~-|\xi|\frac{\Omega\gamma^2}{2d}Im\widetilde{M}t]
\end{array}
\eeq
So from (\ref{40}) we get the decay parameter as
\beq\label{41}
\Gamma=|\xi|\frac{\Omega\gamma^2}{4d}Im\widetilde{M}
\eeq
Now following (\ref{3}), (\ref{4}), (\ref{5}), and (\ref{6}), we get the coherence time as
\beq\label{42}
\tau_C=\frac{2d}{|\xi|\Omega\gamma^2 Im\widetilde{M}}
\eeq
From (\ref{42}), we can easily get that sustainable coherence can occur under the condition
\beq\label{43}
Im\widetilde{M}=0
\eeq
Using equation (\ref{30}) and (\ref{30a}), we get this particular condition as
\beq\label{44}
\tan\phi=\frac{\widetilde{\Delta}\delta_M}{\zeta}
\eeq
where
\beq\label{45}
\mathbf{\zeta=\frac{1}{2}(1-\widetilde{\Delta}^2)\left[|M(\omega_L+\Omega')|-|M(\omega_L)|\right]-|M(\omega_L+\Omega')|}
\eeq
From (\ref{44}) and (\ref{45}), we can see that the condition for sustainable depends on detuning frequency and a few squeezing parameters, which can be controlled in an experimental situation. Let us now consider a special case of squeezing phase \cite{28} of the form of
\beq\label{46}
\phi(\Delta)=\phi(\omega_A)+\frac{\pi\Delta}{\Omega}
\eeq
The squeezing part depending on atomic transition frequency ($\omega_A$) can be taken as $\phi(\omega_A)=0$. Further considering that the deviation from natural atomic transition frequency to be small enough compared to the rabi frequency ($\Delta\ll\Omega$), so that $\tan\phi\approx\phi$, we get from (\ref{44})
\beq\label{47}
\widetilde{\Omega}=\frac{\pi|\zeta|}{\delta_M}
\eeq
This is the condition for sustainable coherence for the particular case of squeezing frequency that we have taken in (\ref{46}).\\

\section{Conclusion}
In this work, we have basically discussed two separate method to sustain the quantum coherence of two-state system in a squeezed vacuum. The first one is more of a principle, where we have asserted that quantum coherence can be sustained by the process of repetitive or quasi-continues measurement. Here measurement means interaction with the squeezing field. This is related to the dynamics of Quantum Zeno effect. In this article, we have taken the ratio of coherence time and dwell time as a quantitative measure of coherence. This ratio gives us the measure of coherence in the period of tunneling through the barrier. If the ratio is greater than unity, then we can assart that the coherence of the system is preserved at least for the period of dwelling, or more. As the ratio goes higher, the coherence is more and more sustainable. With this assertion, we are able to show theoretically, that Zeno dynamics can asymptotically preserve quantum coherence. In the second part, we have derived the the coherence time in terms of various squeezing parameters. Which shows that by controlling those parameters in an experimental situation, we can control the timescale of coherence. We have even been able to derive a specific condition, under which coherence is sustainable. So we can conclude that due to the presence of various non-classical properties \cite{29}, squeezed vacuums are susceptible to procure a coherence preserving environment and can be considered to be a potential candidate for the construction of quantum memory devices.

\end{document}